\documentclass[apj]{emulateapj}
\usepackage{color}
\usepackage{epstopdf}
\usepackage{graphicx}
\usepackage{enumitem}
\def\apj{{ApJ}}
\def\apjl{{ApJL}}
\def\mnras{{ MNRAS}}

\def\nat{{Natur}}
\def\GCN{{ GCN Circ}}

\def\Swift{{\it Swift~}}
\def\actaa{Acta Astron}
\def\nar{New Astronomy Reviews}

\begin{document}
\title{Possible correlations between the emission properties of short GRBs and their offsets from the host galaxies}
\author{Shuai Zhang$^{1,2,3}$, Zhi-Ping Jin$^{1,5}$, Fu-Wen Zhang$^{4}$, Xiang Li$^{1}$, Yi-Zhong Fan$^{1,5}$, Da-Ming Wei$^{1,5}$}
\affil{$^1$ Key Laboratory of Dark Matter and Space Astronomy, Purple Mountain Observatory, Chinese Academy of Sciences, Nanjing, 210008, China.}
\affil{$^2$ University of Chinese Academy of Sciences, Yuquan Road 19, Beijing, 100049, China.}
\affil{$^3$ Department of Space Sciences and Astronomy, Hebei Normal University, Shijiazhuang 050024, China.}

\affil{$^4$ College of Science, Guilin University of Technology, Guilin 541004, China.}
\affil{$^5$ School of Astronomy and Space Science, University of Science and Technology of China, Hefei, Anhui 230026, China}
\email{jin@pmo.ac.cn (ZPJ), yzfan@pmo.ac.cn (YZF), dmwei@pmo.ac.cn (DMW)}

\begin{abstract}
 Short Gamma-Ray Bursts(SGRBs) are widely believed to be from mergers of binary compact objects involving at least one neutron star and hence have a broad range of spatial offsets from their host galaxies. In this work we search for possible correlations between the emission properties of 18 SGRBs and their offsets from the host galaxies. The SGRBs with and without extended emission do not show significant difference between their offset distribution, in agreement with some previous works. There are however possible correlations between the optical and X-ray afterglow emission and the offsets. The underlying physical origins are examined.

\end{abstract}

\section{Introduction}
Gamma-ray Bursts (GRBs), the most violent explosion after the Big Bang, are usually divided into two basic categories: the short GRBs (SGRBs) with a duration shorter than 2 seconds and the long GRBs (LGRBs) that last longer \citep{Kouveliotou1993}. The LGRBs are most-likely powered by the collapse of (rapidly-rotating) massive stars \citep{MacFadyen1999} while the SGRBs likely arise from the coalescence of compact object binaries involving at least one neutron star \citep{Eichler1989, Paczynski1991, Narayan1992}, though the mergers may also produce some long-duration GRBs, too \citep{Gehrels2006,Gal-Yam2006,Fynbo2006,Della Valle2006,Zhang2007,Jin2015,Yang2015}. The smoking-gun signature of the collapsar origin of most LGRBs is the luminous supernovae in the late afterglow emission. The merger of compact object binaries are known to be the promising gravitational wave (GW) sources in the aLIGO/AdVirgo era. The direct observational evidence for the compact object merger origin of SGRBs is still un-available since no GW emission associated with SGRBs has been detected, yet. Before the establishment of SGRB/GW association likely in 2020s \citep{Li2016} when the detection rate will reach to about 1 $ \rm yr^{-1} $, the most important evidence for the merger-origin of some GRBs is the identification of Li-Paczynski macronovae in GRB 130603B, GRB 060614 and GRB 050709 \citep{Tanvir2013, Berger2013ApJL, Yang2015, Jin2016}. If these  macronovae were powered by the NS-BH mergers, the detection prospect by the aLIGO/AdVirgo network is quite promising \citep{Li2016arXiv}. 

The study of the properties of host galaxies of GRBs became feasible after the launch of the BeppoSAX satellite \citep{Boella1997A&AS} which localized the burst accurately. \citet{Wainwright2007} studied the morphological properties of GRB host galaxies and showed that most galaxies have approximately exponential profiles and some are merging and interacting systems. \citet{Bloom2002} studied the locations of LGRBs relative to their host galaxies and found a strong connection between the LGRB location with the star formation region. They also found that the observed offset distribution of LGRBs is consistent with the expected distribution of massive stars in exponential disks (see also \citet{Blanchard2016}). All these findings are in agreement with the collapsar model for long GRBs. For SGRBs, it is more complicated to associate them with their host galaxies due to the faintness of the afterglows and that the binary systems could have traveled far away from their birth sites before the coalescences \citep{Lipunov1997,Fryer1997,Bloom1999,Belczynski2006,Wang2006}. Researches on the host galaxies of SGRBs were not available until 2005 when the SGRB afterglows had been finally discovered \citep{Hjorth2005,Fox2005,Covino2006}. In a study of the spatial offsets of SGRBs from their host galaxies, \citet{Troja2008} showed that among SGRBs, those with extended hard X-ray emission components have small projected physical offsets than those without extended emissions, possibly due to a systematic difference in the progenitors (i.e., the BH-NS and NS-NS mergers give rise to different ``types" of events). If correct, such a finding has far reaching implication on the GW detection \citep{Li2016}. Later, the detailed investigation of {\it Hubble Space Telescope (HST)} observations of SGRB host galaxies \citep{Fong2010, Fong2013} determined the host morphological properties and measured precise physical and host-normalized offsets of SGRBs relative to the galaxy centers. They found that most SGRB hosts are late-type galaxies which have exponential disk profiles and with a median size that is twice as large as that of long GRB hosts. Analysis of the distribution of SGRBs offsets relative to their host galaxy centers indicated that SGRB progenitors are compact object binaries (NS-NS/NS-BH). \citet{Berger2011} also got the same conclusion and ruled out a dominant population of SGRBs from magnetar giant flares. Recently, \citet{LiYe2016} made a detailed comparative study of long and short GRBs in particular on the properties of the host galaxies and the offsets.

Motivated by these previous remarkable progresses, in this work  we search for possible correlations between the emission properties of SGRBs and their offsets from the host galaxies.
In Sec.\ref{sec:data} we describe our data sample. In Sec.\ref{sec:staR} we present the statistical results, discuss the uncertainties and present some preliminary explanation. At last, in Sec. \ref{sec:CanD} we make summary with some discussions.

\section{Data}\label{sec:data}

\citet{Fong2010} published the offsets of nine SGRBs (including angular offset $R_{\rm \theta}$, physical offset $ R_{\rm phy} $, and host-normalized  offset $R_{\rm nor}$) from their host galaxy centers. Later \citet{Fong2013} provided another sample including 16 SGRBs. Among these 25 SGRBs there are just 16 SGRBs with measured redshifts. Plus GRB 060614 \footnote{We should note that GRB 060614 has a longer duration than 2s. However, some of the properties make it more like a  SGRB \citep{Gehrels2006,Gal-Yam2006,Zhang2007}.} from \citet{Blanchard2016} and GRB 150101B from \citet{Fong2016}, we have a sample consisting of 18 SGRBs. For these bursts we collect the total isotropic energy of prompt gamma-ray emission ($ E_{\rm iso}$) and calculate the X-ray ($0.3-10$ keV) afterglow fluence ($F_{\rm X,11}$) at $ 11 \times (1+z) $ hrs post-burst  and optical flux density ($F_{\rm opt,6}$) at $ 6 \times (1+z) $ hrs post-burst , during which the X-Ray and optical afterglow lightcurves were fitted by power law or broken power law model. The optical afterglow data are adopted from \citet{Fong2015} and X-ray data are taken from \citet{Berger2014}, \citet{Fox2005} and \Swift official website $http://www.swift.ac.uk/xrt\_curves/$. Due to the absence of enough afterglow data in a few events in total we have just 13 sets of $F_{\rm X,11}$ and 14 of $F_{\rm opt,6}$. Among those SGRBs, some have $F_{\rm X,11}$ but no $F_{\rm opt,6}$ and some have $F_{\rm opt,6}$ but no $F_{\rm X,11}$. At last, we also collect 13 $ E_{\rm iso}$ \citep{Zhang2012, Zhang2015GCN} out of the total 18 SGRBs. With the $T_{90}$ and redshifts we convert  $ E_{\rm iso}$, $F_{\rm X,11}$ and $F_{\rm opt,6}$  into time-averaged gamma-ray luminosity ($L_{\gamma}$), X-ray luminosity ($L_{\rm X,11}$) and optical luminosity ($L_{\rm opt,6}$). In principle, it is easier to measure the redshifts of the bursts with brighter afterglow emission and this could be a source of the selection effect. However, for a good fraction of SGRBs, the redshifts are not determined with the afterglow spectrum measurements. Instead they are given by the association probability evaluation since most SGRBs are found to be outside of their host galaxies. We therefore suggest that the redshift selected sample may be not seriously biased.

In addition, in order to examine whether there is indeed the difference of host galaxies between SGRBs with and without extended emission\footnote{In this paper, all the extended emission refers to the soft-gamma/hard-X emission following the initial spike. Sometimes the extended emission can not be evident in the gamma-ray band but can be distinguished in the X-ray band.} \citep[EE, see][]{Norris2006}, we also compare the properties of these two sub-groups of SGRBs.

For comparison, we select sample of LGRBs as well. \citet{Blanchard2016} studied 105 LGRBs. From which we select 22 LGRBs with measured redshifts and their X-ray and optical afterglow observations are good enough to obtain reliable $F_{\rm X,11}$ and $F_{\rm opt,6}$. We do the same analysis for LGRBs just like SGRBs and get the set of parameters including angular offset $ R_{\rm \theta} $, physical offset $ R_{\rm phy} $, host-normalized offsets $ R_{\rm nor} $, average isotropic luminosity ($L_{\gamma}$), X-ray and optical afterglow isotropic luminosity ($L_{\rm X,11}$  and $L_{\rm opt,6}$).

In our LGRBs sample, there are two ultra-long GRBs (GRB 060218 and GRB 130925). The duration ($T_{90}$) of GRB 060218 and GRB 130925 are $\sim2100 $s \citep{Campana2006} and $\sim7000 $s \citep{Greiner2014} respectively. It should be noted that GRB 060218 is also a low-luminosity GRB.

\section{Statistical Results }\label{sec:staR}
\subsection{Offset Distribution of SGRBs with and without Extended Emission}

\begin{figure*}
	\begin{center}
		\includegraphics[width=513.5 pt]{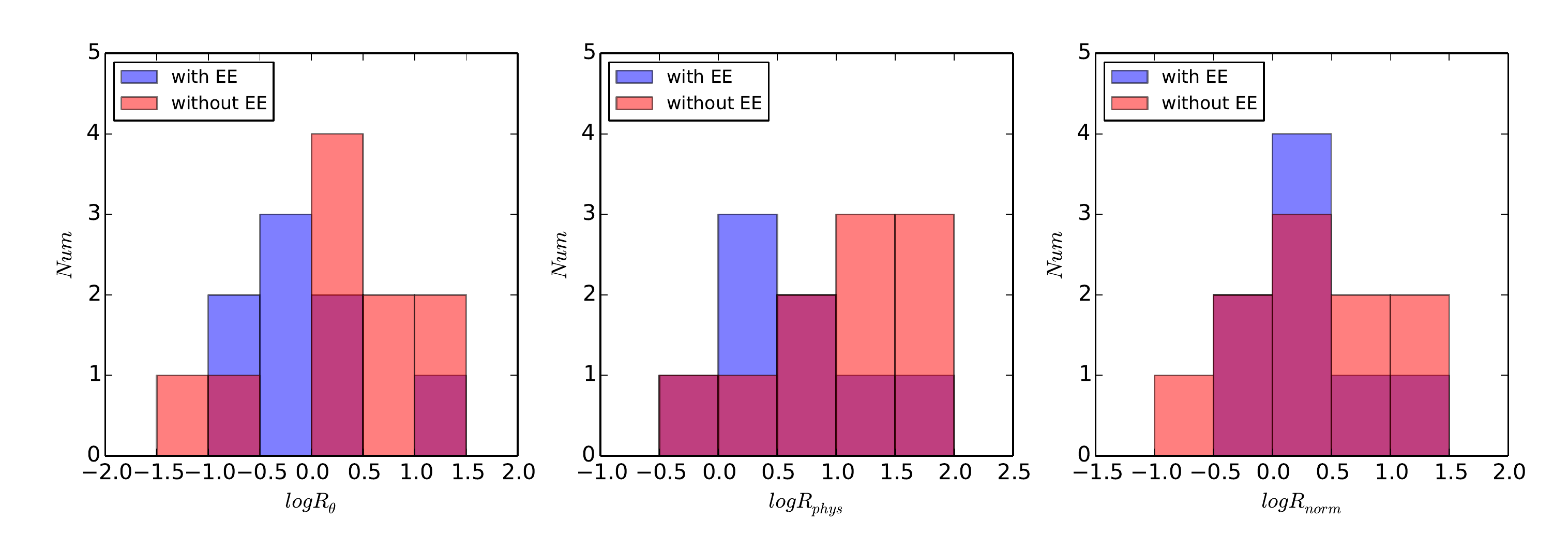}
	\end{center}
	\caption{Comparison of offset distributions of SGRBs with and without extended emission. Blue and red bars represent SGRBs with and without extended emission, respectively. Purple is overlap part of blue and red bars.}
	\label{fig:disoff}
\end{figure*}

In this subsection following \citet{Troja2008} we check whether there is difference between offset distributions of SGRBs with and without EE. Here we also considered X-ray EE while \citet{Troja2008} only considered $ \gamma $-ray EE. Among the 18 SGRBs, 5 have both $\gamma$-ray and X-ray EE and another 3 have just X-ray EE. For SGRBs with and without EE, the distributions of their offsets from host galaxies are shown in Fig. \ref{fig:disoff}. The offsets of SGRBs with and without EE seem to have similar distribution. In order to verify this we take a K-S test to examine the relation of the two distributions and find the p-values are (0.30, 0.24, 0.83) for ($R_{\rm \theta} $, $R_{\rm phy}$, $R_{\rm nor}$), respectively. This indicates that there is no significant difference between the two subsample distributions, consistent with \citet{Fong2010}, \citet{Salvaterra2010} and \citet{Fong2013}.

\subsection{Correlation Between Luminosities and their Offsets From the Host Galaxy Centers}

\begin{figure*}
	\begin{center}
		\includegraphics[width = 513.5pt]{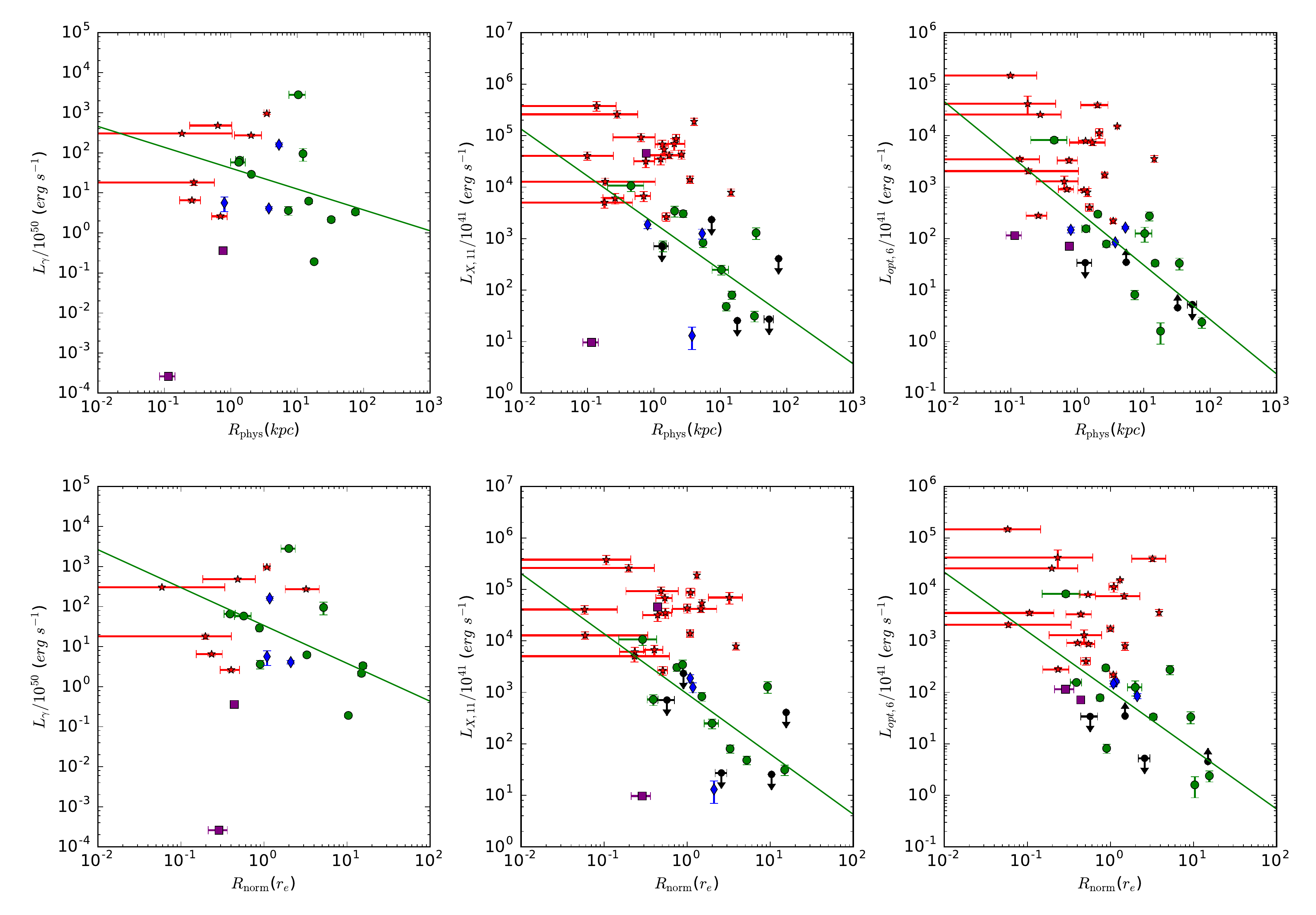}
		\caption{Correlations between luminosities of GRBs and their offsets from host galaxy centers. The red stars, purple square points and green circular points are for long GRBs, ultra-long GRBs and SGRB respectively. The Blue diamond points represent SGRBs with macronova signals. The solid green line in each panel is the fit to all SGRB sample. Arrows represent upper or lower limits of SGRBs. }
		\label{fig:correlation}
	\end{center}
	
\end{figure*}

Now we turn to search for possible correlations between the GRB/afterglow luminosities (i.e., $L_{\gamma}$, $L_{\rm X,11}$  and $L_{\rm opt,6}$) and the offsets ($R_{\rm phy}$, $R_{\rm nor}$). In Fig.\ref{fig:correlation} we show the data of all SGRBs. The blue diamond points and the green circular points represent SGRBs with and without macronova signals. There is a clear trend that the farther the SGRB is from the center of the host galaxy, the lower the afterglow (both in X-ray and optical bands) luminosity. But the average isotropic prompt (gamma-ray band) luminosity dose not follow this trend. To obtain the quantitative relationship, we use power law model to fit the data of all SGRB sample. The green lines are the best fit results and their expressions and correlation coefficients are summarized in Tabe \ref{tab:fit}. The most significant correlations are $L_{\rm X,11} \propto R_{\rm norm}^{-1.16 \pm 0.57}$ and  $L_{\rm opt,6} \propto R_{\rm norm}^{-1.23 \pm 0.51}$ and their correlation coefficients are 0.66 and 0.70 , respectively.

For comparison, we also draw LGRBs and ultra-long GRBs in each panel of Fig.\ref{fig:correlation}, LGRBs generally located at the upper left side of SGRBs in the two-dimensional map and their luminosities are independent of offsets. All the absolute values of correlation coefficients are smaller than 0.33. Due to the limited number of ultra-long GRB samples, no general conclusion on their distributions can be drawn (note that GRB 060218 is distinguished for its very low luminosity and small offset). 

There are several observational biases which might impact on the observed correlations. For instance, it is easier to detect bright optical afterglows close to the center of galaxies than faint ones, which may soften the luminosity-offset correlation. However, we think the effect is not significant for the following reasons. First, it should be easy to detect bright afterglows with large offsets, but there is no burst on the upper-right side of each panel in Fig.\ref{fig:correlation}. 	
Second, if there was a SGRB close to the center of the host and its afterglow was faint (it lied on the lower-left side of the correlations), it might be detected in the X-ray band because the X-ray afterglows are less affected by the host galaxies \citep{Le2003} and have a much higher detection rate\citep{Fong2015}. In some cases (GRB050509B, GRB061201, GRB070724A, GRB100117A), they are well located but only have upper or lower limits of $ L_{\rm opt,6} $. All these events have large offsets and still potentially agree with the observed correlations, see Fig.\ref{fig:correlation}. For other cases with detection in X-ray but not in optical (\citet{Rossi2012}, they are excluded in our sample because their host galaxies had not been observed by HST), usually their positions are not well determined, the centers of some host galaxies are enclosed in the X-ray error. If these events were indeed from the inner host galaxies, although the extinctions of SGRBs are usually not  significant, there is still a possibility that the non-detection of optical afterglow may be due to the extinction \citep{Rossi2012} and the intrinsic optical emission could still follow the correlation.
Finally, if this bias is significant, then it is same for both short and long GRB samples, but no similar correlation is found in LGRBs.

What's more, the image subtraction techniques would weaken this effect. In the near future these correlations can be tested by ALMA which will look deeper into the inside of host galaxies of SGRBs.

\begin{table}
	\caption{fitting of the sgrb{\rm s} sample.}
\begin{tabular}{|l|c|c|}
	\hline Correlation & $R$\footnotemark[1] \\
	\hline $logL_{\rm \gamma}=(-0.52 \pm 0.49) logR_{\rm phys} +(51.61 \pm 0.48)$ & -0.29 \\
	\hline $logL_{\rm \gamma}=(-0.94 \pm 0.54) logR_{\rm norm} +(51.52 \pm 0.33)$ & -0.48 \\
	\hline $logL_{\rm X, 11}=(-0.91 \pm 0.49) logR_{\rm phys} +(44.30 \pm 0.43)$ & -0.60 \\
	\hline $logL_{\rm X, 11}=(-1.16 \pm 0.57) logR_{\rm norm} +(43.96 \pm 0.31)$ & -0.66 \\
	\hline $logL_{\rm opt, 6}=(-1.12 \pm 0.44) logR_{\rm phys} +(43.69 \pm 0.42)$ & -0.76 \\
	\hline $logL_{\rm opt, 6}=(-1.23 \pm 0.51) logR_{\rm norm} +(43.19 \pm 0.30)$ & -0.70 \\
	\hline
\end{tabular}\label{tab:fit}
\footnotetext[1]{Correlation Coefficient between luminosity and offset.}
\end{table}

\subsection{Discussion}

The offset distribution of NS-NS binaries in Milky Way type galaxies has been widely examined \citep{Lipunov1997,Bloom1999,Fryer1999,Belczynski2006}, which is found to be consistent with the  observed offset distribution of SGRBs \citep{Berger2011,Fong2010,Fong2013} though the possibility of the existence of other progenitor systems can not be ruled out. Such a result has been taken as one piece of compelling evidence for the compact object merger origin model. As shown in Fig.2, more than 60\% LGRBs have been found inside their host galaxies\citep{Blanchard2016,Lyman2017} while about 70\% of SGRBs located outside their host galaxies.

The prompt gamma-ray emission of GRBs has been widely attributed to the internal energy dissipation of the unsteady outflow material launched by the central engines \citep{Kumar2015}. Hence the prompt emission luminosities are mainly governed by the central engines and should not display significant dependence on the offsets from the host galaxies. One exception is that the SGRB progenitors might mainly consist of two sub-groups, one is the NS-BH binaries which may obtain smaller kick velocities for their lager system mass and the other is the NS-NS binaries. The former may be more concentrated around the host galaxy center while the latter may have a larger typical offset. If the NS-BH merger driven GRBs are more typical than the NS-NS merger origin GRBs, one may expect some dependence of the prompt emission luminosities on the offset, as argued in \citet{Troja2008}. The current prompt emission data presented in Fig.2 are still insufficient to draw a reliable conclusion.

The presence of possible correlations between the X-ray and optical afterglow emission and the offsets are somewhat ``expected". This is because the number density of circum-burst medium ($n_{\rm e}$) is expected to decrease with the distance to the host galaxy center \citep{Fong2015} and the afterglow emission flux $F$ depends on $n_{\rm e}$ as long as the observer's frequency $\nu_{\rm obs}$ is $<\max\{\nu_{\rm c},\nu_{\rm m}\}$ ( where the $ \nu_m $ and $ \nu_c $ refers to typical synchrotron frequency and cooling frequency \citep{Sari1998} respectively. Note that in the standard afterglow model, for $\nu_{\rm obs}>\max\{\nu_{\rm c},\nu_{\rm m}\}$ we have the afterglow flux independent of $n_{\rm e}$). For example, in the most-likely scenario of $\nu_{\rm m}<\nu_{\rm obs}<\nu_{\rm c}$ we have $F\propto n_{\rm e}^{1/2}$. Supposing that beyond a distance from the center there is a relation $n_{\rm e}\propto R^{-\alpha}$, we will have $F\propto R^{-\alpha/2}$. If the observed dependence of $F$ on the offset is solely due to such an effect, we can constrain $\alpha \sim 2.2$. So far it is unclear whether it is the case or not. The kinetic energy($ E_{\rm k} $) also have an impact on $F$, but it seems to be independent of the offset from host galaxy center. 

As a comparison, LGRBs from massive star collapse occurred in star forming region of galaxies and their progenitors are almost static to their birthplace \citep{Bloom2002}. This model is supported  by the non-correlation between long GRB/afterglow luminosities and their offsets from host galaxies.

\section{Summary}\label{sec:CanD}
In this work we carry out some statical analysis focusing on the possible dependence of the emission properties on the offsets of SGRBs. We find possible correlations between short GRB afterglow luminosities and their offsets from host galaxies that can be expressed as
$L_{\rm X,11} \propto R_{\rm norm}^{-1.16 \pm 0.57}$
and  $L_{\rm opt,6} \propto R_{\rm norm}^{-1.23 \pm 0.51}$ (please see Table 1 for the correlation cofficients).
These correlations are somewhat ``expected" since the number density of circum-burst medium ($n_{\rm e}$) should decrease with the distance to the host galaxy center (i.e., $R$) and the afterglow emission flux $F$ depends on $n_{\rm e}$ as long as the observer's frequency $\nu_{\rm obs}<\max\{\nu_{\rm c},\nu_{\rm m}\}$ (For example, in the scenario of $\nu_{\rm m}<\nu_{\rm obs}<\nu_{\rm c}$ we have $F\propto n_{\rm e}^{1/2}$). Hence our result may have shed some lights on the dependence of $n_{\rm e}$ on $R$.

Besides, there are some other uncertainties related to the correlation. For instance, the angle between the line of sight and the host galaxy disk may effect the measured offset although the affection may be weakened in the statistical study. 
Another uncertainty may be the SGRBs occurred in globular cluster. The natal kick velocities and the merge rate of binaries in globular cluster may be affected by the total globular cluster system. \citet{Grindlay2006} presented numerical result that SGRBs from binary neutron star mergers in globular clusters account for $ \sim 10-30 \% $ of total. The influence exerted by SGRBs occurred in globular cluster still need more further discussion.

\section*{Acknowledgments}

This work was supported in part by 973 Programme of China under grants (No. 2014CB845800 and No. 2013CB837000), National Natural Science Foundation of China under grants 11525313, 11433009, and 11303098, the Chinese Academy of Sciences via the Strategic Priority Research Program (No. XDB23040000) and the External Cooperation Program of BIC (No. 114332KYSB20160007).

\end{document}